\documentclass[12pt]{article}
\usepackage{amsfonts,amssymb,amsmath}
\usepackage[dvips]{epsfig}
\begin{document}
\title{\bf Perfect transfer of coherent state-based qubits via coupled cavities}
\author{ N. Behzadi $^{a}$
\thanks{E-mail:n.behzadi@tabrizu.ac.ir}  ,
S. Kazemi Rudsary$^{b}$  ,
B. Ahansaz Salmasi$^{b}$
\\ $^a${\small Research Institute for Fundamental Sciences,}
\\ $^b${\small Department of Theoretical Physics
and Astrophysics, Faculty of Physics,}
\\ {\small University of
Tabriz, Tabriz 51666-16471, Iran.}} \maketitle
\begin{abstract}
\noindent Motivated by the need for communication of coherent state-based qubits in quantum computers, we introduce a method for perfect transferring of an arbitrary superposition of coherent states between two distant nodes of a linear array of three semiconductor QDs. The QDs trapped in a system of coupled cavities. In this method, the field mode of the cavities, as the resource of transferring of quantum states, are only virtually excited which minimizes the effect of decoherence due to photon loss.
\end{abstract}

\section{Introduction}
The task of coherently transferring quantum states through distant
parties has been of utmost importance in the field of quantum
information and computation. Cavity quantum electrodynamics (QED)
system with atoms or quantum dots (QDs) as artificial atoms
interacting with quantized electromagnetic fields are almost an
ideal candidate for implementing quantum information processing
protocols. Because atoms are suitable for storing information
and photons are suitable for transporting it \cite{cirac1}. Indeed, the system of high-Q cavities and atoms in the strong interaction regime is one of experimentally realizable systems in which
the intrinsic quantum mechanical coupling dominates losses that due to dissipation \cite{cirac1, hen, khi}. To accomplish the task of quantum state transfer, several approaches have been employed using cavity QED system. For example, communications of
quantum states over long distances, cavity QED system are
used to transfer quantum states from a given atom to another through
photons transmitting in an optical fiber \cite{cirac1}. While for short
distance quantum communication, exploiting the natural dynamics of
the many-body system, as was first introduced by Bose in the form of
spin chain \cite{bose}, has become evident \cite{irish, peng}. In spin chains, single spin
addressing is difficult because the spatial separation between
neighboring spins is very small \cite{peng}. Thus the control over the
couplings between the spins or over individual spins is very hard to
achieve. While the coupled cavities, which each of them contains an atom, have the advantage of easily addressing
individual cavities with optical lasers. Furthermore, the
interaction of a cavity and a QD, as an artificial atom, can be engineered in such way that
the QD trapped in the cavity can have relatively long-lived energy
levels which is suitable for encoding of quantum information \cite{yabu}.

On the other hand, quantum computation on the basis of superposition
of coherent states as logical qubits, coherent state-based qubits, has several advantages
\cite{glan, jeo, ral}. Therefore, transferring of quantum information encoded on a cetain
superposition of coherent states, as a qubit, can be thought as an unavoidable
task. Since semiconductor
QDs are appealing for realization of quantum computer and quantum
information processing \cite{sander, yu}, then preparing a typical
superposition of coherent states in a QD and transferring it to another, using a system of coupled cavities, can
be regarded as an efficient way for communication of coherent state-based qubits in a quantum computer.

In this work we propose a protocol for transferring an arbitrary
coherent state-based qubit through a linear array of three single-mode
coupled cavities. The system under consideration contains three QDs trapped in a system of cavity QED. We show that, by suitable setting of parameters, the dynamical evolution of the system can transfer an arbitrary coherent state-based qubit
prepared on excitonic mode of QD at one end of the array to excitonic mode of QD on the other end with perfect fidelity. The process of perfect transferring of a quantum state, in this way, can be established in two distinct regimes. In one of them, the field mode of the cavities is in resonance interaction with the excitonic mode of the QDs. Therefore, the field mode can be extremely populated, which in turn, leads to decay of photons and therefore lose of coherency of the system \cite{Haroche, Piza}. But, when the field mode is highly detuned with the excitonic mode, the establishment of perfect state transfer can be satisfied without populating the field mode in the cavities. In this situation, the perfect state transfer protocol is susceptible to decoherence
from photon loss and thus, during the operation, the efficient decoherence rate of the cavities
is greatly prolonged.

This paper is laid out in the following way. In section 2, we start to describe the basic properties of the scheme. In Section 3, we describe realization of the scheme and representation a process for dominating on the decoherence influencing the system to lose its coherency.  The paper is ended with a brief conclusion.

\section{ Hamiltonian description}
The system under consideration contains three quantum dots trapped
in a linear array of three separated single-mode
coupled cavities as shown in Fig. 1. The conditions for achievement of perfect transfer of a typical superposition of standard coherent states are that the Bohr radius $a_{B}$ of the bulk exciton is so small in comparison to the effective radius R of each quantum dot. It is
assumed that there are only a few electrons excited from valanced-band to
conduction-band. This leads to the fact that, for an effective region with
excitonic Bohr radius, the average
number of excitons is no more than one in that region. Consequently, by these assumptions, excitons can be approximately considered as bosons. Also, exciton-exciton interactions and phase-space filling
effect as nonlinear terms can be neglected and the ground energy
of the excitons in each quantum dot is taken to be the same. We consider
frequency of field mode in cavities by $\omega_{c}\equiv\omega$ and frequency of
excitonic mode by $\omega_{e}\equiv\omega-\Delta$ where $\Delta$ is the
detuning between the excitonic mode and the field mode.
The Hamiltonian under the rotating wave approximation is given by
\begin{eqnarray}
&&\hspace{-7mm}\hat{H}=\hbar \omega
\sum_{i=1}^{3}\hat{a}_{i}^{\dagger}\hat{a}_{i}+\hbar (\omega-\Delta)
\sum_{i=1}^{3}\hat{b}_{i}^{\dagger}\hat{b}_{i}+
g\sum_{i=1}^{3}\left(\hat{b}_{i}^{\dagger}\hat{a}_{i}+\hat{b}_{i}\hat{a}_{i}^{\dagger}\right)
+\hbar
c\sum_{i=1}^{2}\left(\hat{a}_{i}^{\dagger}\hat{a}_{i+1}+\hat{a}_{i+1}\hat{a}_{i}^{\dagger}\right),
\end{eqnarray}
where $\hat{a}_{i}^{\dagger}$ and $\hat{a}_{i}$ are creation and
annihilation operators for the field mode in the ith cavity and
$\hat{b}_{i}^{\dagger}$ and $\hat{b}_{i}$ are creation and
annihilation operators for the excitonic mode in the ith QD.
$g$ is the coupling constant between each quantum dot and the
relative cavity field and $c$ is the coupling strength between the
cavities whose magnitude provides amount of photon hopping
between the cavities. Indeed, photon hopping occurs due to overlap of the field modes in adjacent cavities \cite{hart}
,which in turn, arisen from tunneling of the field modes
between adjacent cavities \cite{huo}. The dynamical problem is
completely specified by the following set of coupled equations for
operators of the field modes and excitons as
\begin{eqnarray}
\dot{\hat{a}}_{1}=-i\omega \hat{a}_{1}-ig \hat{b}_{1}-ic
\hat{a}_{2},
\end{eqnarray}
\begin{eqnarray}
\dot{\hat{a}}_{2}=-i\omega \hat{a}_{2}-ig \hat{b}_{2}-ic
\hat{a}_{1}-ic \hat{a}_{3},
\end{eqnarray}
\begin{eqnarray}
\dot{\hat{a}}_{3}=-i\omega \hat{a}_{3}-ig \hat{b}_{3}-ic
\hat{a}_{2},
\end{eqnarray}
\begin{eqnarray}
\dot{\hat{b}}_{j}=-i(\omega-\Delta) \hat{b}_{j}-ig
\hat{b}_{j},\quad\ j=1, 2, 3.
\end{eqnarray}
This linear system of first order differential equations can be
easily solved, obtaining
\begin{eqnarray}
\hat{b}_{1}(t)=u_{_{b_{1}1}}(t)\hat{a}_{1}(0)+u_{_{b_{1}2}}(t)\hat{b}_{1}(0)+u_{_{b_{1}3}}(t)\hat{a}_{2}(0)+u_{_{b_{1}4}}(t)\hat{b}_{2}(0)+u_{_{b_{1}5}}(t)\hat{a}_{3}(0)+u_{_{b_{1}6}}(t)\hat{b}_{3}(0),
\end{eqnarray}
where $u_{_{b_{1}j}}(t)$s for $j=1,2,...,6$ are as
\begin{eqnarray}
u_{_{b_{1}1}}(t)=-ie^{-i(\omega-\Delta/2)t}\{e^{-ict/\sqrt{2}}\frac{g}{2A}sin(At/2)+e^{ict/\sqrt{2}}\frac{g}{2B}sin(Bt/2)+\frac{g}{F}sin(Ft/2)\},
\end{eqnarray}
\begin{eqnarray}
\begin{array}{c}
  u_{_{b_{1}2}}(t)=e^{-i(\omega-\Delta/2)t}\{e^{-ict/\sqrt{2}}(\frac{1}{4}cos(At/2)+i\frac{\Delta+c\sqrt{2}}{4A}sin(At/2))
  \\\\
  +e^{ict/\sqrt{2}}(\frac{1}{4}cos(Bt/2)+i\frac{\Delta-c\sqrt{2}}{4B}sin(Bt/2))+\frac{1}{2}cos(Ft/2)+i\frac{\Delta}{2F}sin(Ft/2)\}, \\
\end{array}
\end{eqnarray}
\begin{eqnarray}
u_{_{b_{1}3}}(t)=ie^{-i(\omega-\Delta/2)t}\{e^{-ict/\sqrt{2}}\frac{-g\sqrt{2}}{2A}sin(At/2)+e^{ict/\sqrt{2}}\frac{\sqrt{2}g}{2B}sin(Bt/2)\},
\end{eqnarray}
\begin{eqnarray}
\begin{array}{c}
  u_{_{b_{1}4}}(t)=e^{-i(\omega-\Delta/2)t}\{e^{-ict/\sqrt{2}}(\frac{\sqrt{2}}{4}cos(At/2)+i\frac{2\Delta+c\sqrt{2}}{4A}sin(At/2))
  \\\\
  -e^{ict/\sqrt{2}}(\frac{\sqrt{2}}{4}cos(Bt/2)+i\frac{2\Delta+c\sqrt{2}}{4B}sin(Bt/2))\}, \\
\end{array}
\end{eqnarray}
\begin{eqnarray}
u_{_{b_{1}5}}(t)=-ie^{-i(\omega-\Delta/2)t}\{e^{-ict/\sqrt{2}}\frac{g}{2A}sin(At/2)+e^{ict/\sqrt{2}}\frac{g}{2B}sin(Bt/2)-\frac{g}{F}sin(Ft/2)\},
\end{eqnarray}
\begin{eqnarray}
\begin{array}{c}
  u_{_{b_{1}6}}(t)=e^{-i(\omega-\Delta/2)t}\{e^{-ict/\sqrt{2}}(\frac{1}{4}cos(At/2)+i\frac{\Delta+c\sqrt{2}}{4A}sin(At/2))
  \\\\
  +e^{ict/\sqrt{2}}(\frac{1}{4}cos(Bt/2)+i\frac{\Delta-c\sqrt{2}}{4B}sin(Bt/2))-\frac{1}{2}cos(Ft/2)-i\frac{\Delta}{2F}sin(Ft/2)\}. \\
\end{array}
\end{eqnarray}
In the above expressions,
$A=\sqrt{\Delta^{2}+2c^{2}+4g^{2}+2\sqrt{2}\Delta c}$,
$B=\sqrt{\Delta^{2}+2c^{2}+4g^{2}-2\sqrt{2}\Delta c}$ and
$F=\sqrt{\Delta^{2}+4g^{2}}$. It is easy to show that
\begin{eqnarray}
\sum_{j=1}^{6}|u_{_{b_{1}j}}(t)|^{2}=1,
\end{eqnarray}
which in turn, is consequence of the fact that the number of
excitations, i.e. $\hat{N}=\sum_{i=1}^{3}\hat{a}_{i}^{\dagger}\hat{a}_{i}+
\sum_{i=1}^{3}\hat{b}_{i}^{\dagger}\hat{b}_{i}$, commutes with the
Hamiltonian and therefore it is a constant of the motion. Each of $|u_{_{b_{1}j}}(t)|^{2}$ $(j=1, 2, ..., 6)$ is called probability of excitation or population related to the field mode of cavities or excitonic mode of QDs.

Let us consider two coherent states $|\alpha\rangle$ and $|-\alpha\rangle$ with coherent amplitude $|\alpha|$ and the overlap of them as $\langle-\alpha|\alpha\rangle=e^{-2|\alpha|^{2}}$. We identify the two coherent states as basis states for logical qubit as
\begin{eqnarray}
|0\rangle_{L}\equiv|\alpha\rangle,\quad\ |1\rangle_{L}\equiv|-\alpha\rangle.
\end{eqnarray}
A properly normalized qubit based on superposition of coherent states $|\alpha\rangle$ and $|-\alpha\rangle$ (coherent state-based qubit), is given by
\begin{eqnarray}
|Q(\alpha)\rangle=\frac{1}{\sqrt{N_{\alpha}}}(\mu|\alpha\rangle+\nu|-\alpha\rangle)
\end{eqnarray}
where $N_{\alpha}=|\mu|^{2}+|\nu|^{2}+e^{-2|\alpha|^{2}}(\mu\nu^{\ast}+\mu^{\ast}\nu)$ is the normalization factor, and $\mu$ and $\nu$ are complex numbers. We expect that the scheme proposed as above, could provide transferring the coherent state-based qubit (15), prepared on the excitonic mode of the QD, from one end of the channel of coupled cavities to the other end.
\section{ Perfect transfer of coherent state-based qubit}
We now consider the system to be in an initial state in which the excitonic mode of QD trapped in the first cavity is prepared in a superposition of coherent excitonic states as $|Q(\alpha)\rangle$ in (15) and the other field modes of the cavities and the exciton
modes of QDs are in the respective vacuum states as below
\begin{eqnarray}
\begin{array}{c}
  \left|\psi(0)\right\rangle=\frac{1}{\sqrt{N_{\alpha}}}\left|0\right\rangle_{c_{1}}(\mu|\alpha\rangle+\nu|-\alpha\rangle)_{e_{1}}\left|0\right\rangle_{c_{2}}\left|0\right\rangle_{e_{2}}\left|0\right\rangle_{c_{3}}\left|0\right\rangle_{e_{3}}
 \\\\
  =\frac{1}{\sqrt{N_{\alpha}}}\left|0\right\rangle_{c_{1}}(\mu e^{\alpha \hat{b}^{\dagger}_{1}(0)-\alpha^{\ast} \hat{b}_{1}(0)}+\nu e^{-\alpha \hat{b}^{\dagger}_{1}(0)+\alpha^{\ast} \hat{b}_{1}(0)})\left|0\right\rangle_{e_{1}}\left|0\right\rangle_{c_{2}}\left|0\right\rangle_{e_{2}}\left|0\right\rangle_{c_{3}}\left|0\right\rangle_{e_{3}},
  \end{array}
\end{eqnarray}
where we have used the definition of constructing the standard coherent states as follows
\begin{eqnarray}
|\alpha\rangle_{e_{1}}\equiv e^{\alpha \hat{b}^{\dagger}_{1}(0)-\alpha^{\ast} \hat{b}_{1}(0)}|0\rangle_{e_{1}}.
\end{eqnarray}
The time evolution of the initial state of the system is obtained as
\begin{eqnarray}
&&\hspace{-7.5mm}\left|\psi(t)\right\rangle=\hat{U}(t)\left|\psi(0)\right\rangle
\nonumber
\\
&&\hspace{-7.5mm}=\frac{1}{\sqrt{N_{\alpha}}}(\mu e^{\alpha \hat{b}^{\dagger}_{1}(t)-\alpha^{\ast} \hat{b}_{1}(t)}+\nu e^{-\alpha \hat{b}^{\dagger}_{1}(t)+\alpha^{\ast} \hat{b}_{1}(t)})\left|0\right\rangle_{c_{1}}\left|0\right\rangle_{e_{1}}\left|0\right\rangle_{c_{2}}\left|0\right\rangle_{e_{2}}\left|0\right\rangle_{c_{3}}\left|0\right\rangle_{e_{3}}
,
\end{eqnarray}
where $\hat{U}(t)$ is the time evolution unitary operator. Note that in obtaining the equation (18) we have used the relations: $\hat{U}(t)\hat{O}(0)\hat{U}^{\dagger}(t)=\hat{O}(t)$ and $\hat{U}(t)\left|0\right\rangle=|0\rangle$ with $|0\rangle\equiv\left|0\right\rangle_{c_{1}}\left|0\right\rangle_{e_{1}}\left|0\right\rangle_{c_{2}}\left|0\right\rangle_{e_{2}}\left|0\right\rangle_{c_{3}}\left|0\right\rangle_{e_{3}}
$. We now expect that after a certain time, namely $t^{\ast}$, the coherent state-based qubit prepared in the excitonic mode of the first QD finds itself in the excitonic mode of the third QD, i. e.
\begin{eqnarray}
\left|\psi(t^{\ast})\right\rangle=\frac{1}{\sqrt{N_{\alpha}}}\left|0\right\rangle_{c_{1}}\left|0\right\rangle_{e_{1}}\left|0\right\rangle_{c_{2}}\left|0\right\rangle_{e_{2}}\left|0\right\rangle_{c_{3}}\left(\mu|\alpha\rangle+\nu|-\alpha\rangle\right)_{e_{3}}.
\end{eqnarray}
This, in turn, shows that the superposition of coherent excitonic states prepared in the first QD has been perfectly transferred to the third QD. But is this situation really possible? In general, behavior
of the system is governed by the relative values of the
three parameters g, $\Delta$, and c. In equation (6), the time dependency of the operator $\hat{b}_{1}(t)$ is evident. Therefore, the possibility of satisfying the equation (19) is provided that
\begin{eqnarray}
\hat{b}_{1}(t^{\ast})=\hat{b}_{3}(0).
\end{eqnarray}
By considering the equation (6) and (13), this is possible when, for a certain time $t^{\ast}$, $|u_{_{b_{1}6}}(t^{\ast})|^{2}=1$ which is equivalent to that $|u_{_{b_{1}j}}(t^{\ast})|^{2}=0$ for $j=1, 2, ..., 5$. This situation can be satisfied for various amounts of the parameters g, $\Delta$, and c. For example at resonance case with $c=1$, $g=65$ and $\Delta=0$, Fig. 2 shows that satisfying the statement $|u_{_{b_{1}6}}(t^{\ast})|^{2}=1$, ensures that the perfect transfer of coherent state-based qubit in (15) from the first QD to the third one has been done successfully at a certain time $t^{\ast}$ called transfer time. On the other hand, at the resonance, the field mode of the cavities with the average number of photons calculated by
\begin{eqnarray}
\bar{n}=\sum_{i=1}^{3}\langle\psi(t)|\hat{a}^{\dagger}_{i}\hat{a}_{i}|\psi(t)\rangle=\frac{|\alpha|^{2}\left(|\mu|^{2}+|\nu|^{2}-e^{-2|\alpha|^{2}}(\mu\nu^{\ast}+\mu^{\ast}\nu)\right)F(t)}{|\mu|^{2}+|\nu|^{2}+e^{-2|\alpha|^{2}}(\mu\nu^{\ast}+\mu^{\ast}\nu)},
\end{eqnarray}
are considerably populated. This is evident due to the presence of the factor
\begin{eqnarray}
F(t)=|u_{_{b_{1}1}}(t)|^{2}+|u_{_{b_{1}3}}(t)|^{2}+|u_{_{b_{1}5}}(t)|^{2},
\end{eqnarray}
called probability of populations of the field mode of cavities as shown in Fig. 2. This, in turn, means that the average number of photon in each cavity can be considerably large. But, in the presence of interaction between the system and environment, it is evident from \cite{Haroche, Piza} that: the larger the average number of photons
inside the cavity, the faster will the coherence decay. Therefore, the lose of coherency of the system is unavoided. To remove the effect of decoherence on the system as arisen as above, we should prevent the populating of the field mode of the cavities. This situation can be achieved for some nonzero values of the detuning parameter, i.e. $\Delta\neq0$, as shown in Fig. 3. The detuning parameter $\Delta$ may be chosen in such way that the factor $F(t)$ becomes so small which in turn, leads to the vanishing of the average number of photon in the field mode of cavities. Therefore, during the process of perfect transferring of a typical coherent state-based qunit which takes place slowly, the field mode of the cavities only virtually excited. In the other words, transferring a general quantum state from one end of the channel of coupled cavities to the other one can be done by virtual photons, protecting against decoherence via cavity decay \cite{irish, biao, majer}.

\section{Conclusion}
We have presented a protocol for perfect transfer of coherent state-based qubit. This protocol utilize the cavity field, induced nonlocal interaction, to couple two QDs on the opposite sides of the channel. The interaction is mediated by the exchange of virtual photons rather than real photons, avoiding cavity-induced loss. Therefore, the perfect transferring of quantum information from one end of the channel to the other end without occurring efficient decoherence can be achieved. This, in turn, represents an interesting step toward the realization of quantum communication in quantum computers.

\newpage
\textbf{Figure Captions}
\itemize{}
\item Fig. 1. The system consists of three identical
single-mode optical cavities, and each cavity
contains a QD. The cavities are placed on a linear array and coupled to each other by the overlap of the field modes of the adjacent cavities with
a hopping strength c.
\begin{figure}
\centering
\includegraphics[width=445 pt]{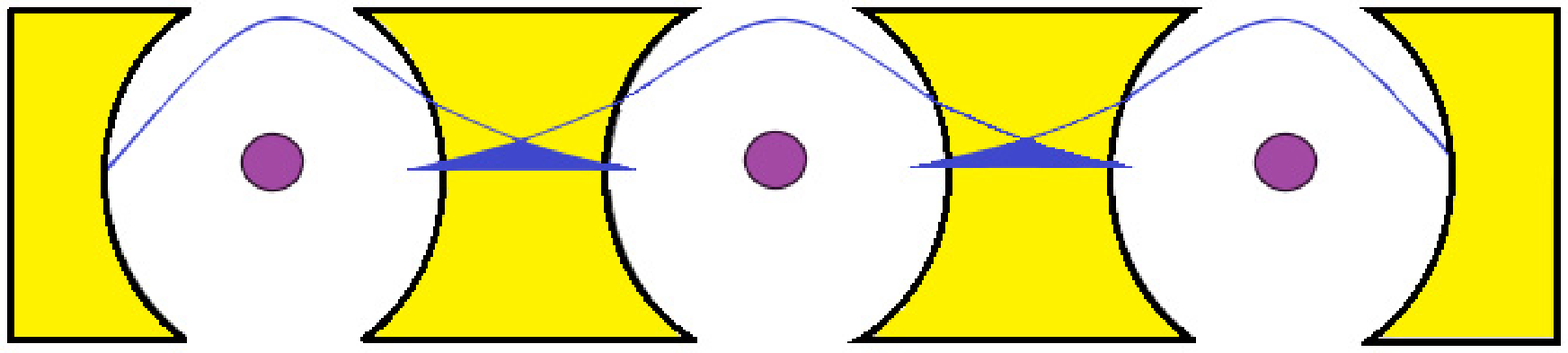}
\caption{} \label{Fig1}
\end{figure}
\newpage
\textbf{Figure Captions}
\itemize{}
\item Fig. 2. Perfect transfer of an arbitrary coherent state-based qubit (15), at the resonance regime with  $c=1$ and $g=65$ (in
units of $\omega$).
 The blue curve represents the population of the field mode of cavities, i.e. $F=|u_{_{b_{1}1}}(t)|^{2}+|u_{_{b_{1}3}}(t)|^{2}+|u_{_{b_{1}5}}(t)|^{2}$. While $U_{2}=|u_{_{b_{1}2}}(t)|^{2}$, $U_{4}=|u_{_{b_{1}4}}(t)|^{2}$ and $U_{6}=|u_{_{b_{1}6}}(t)|^{2}$ represented by yellow, red, and green curves respectively, are the populations of excitonic mode in QDs. The expression $U_{6}=|u_{_{b_{1}6}}(t^{\ast})|^{2}=1$ with transfer time $t^{\ast}=4.4464$, ensures perfect transfer of an arbitrary coherent state-based qubit from first QD to third one.
\begin{figure}
\centering
\includegraphics[width=445 pt]{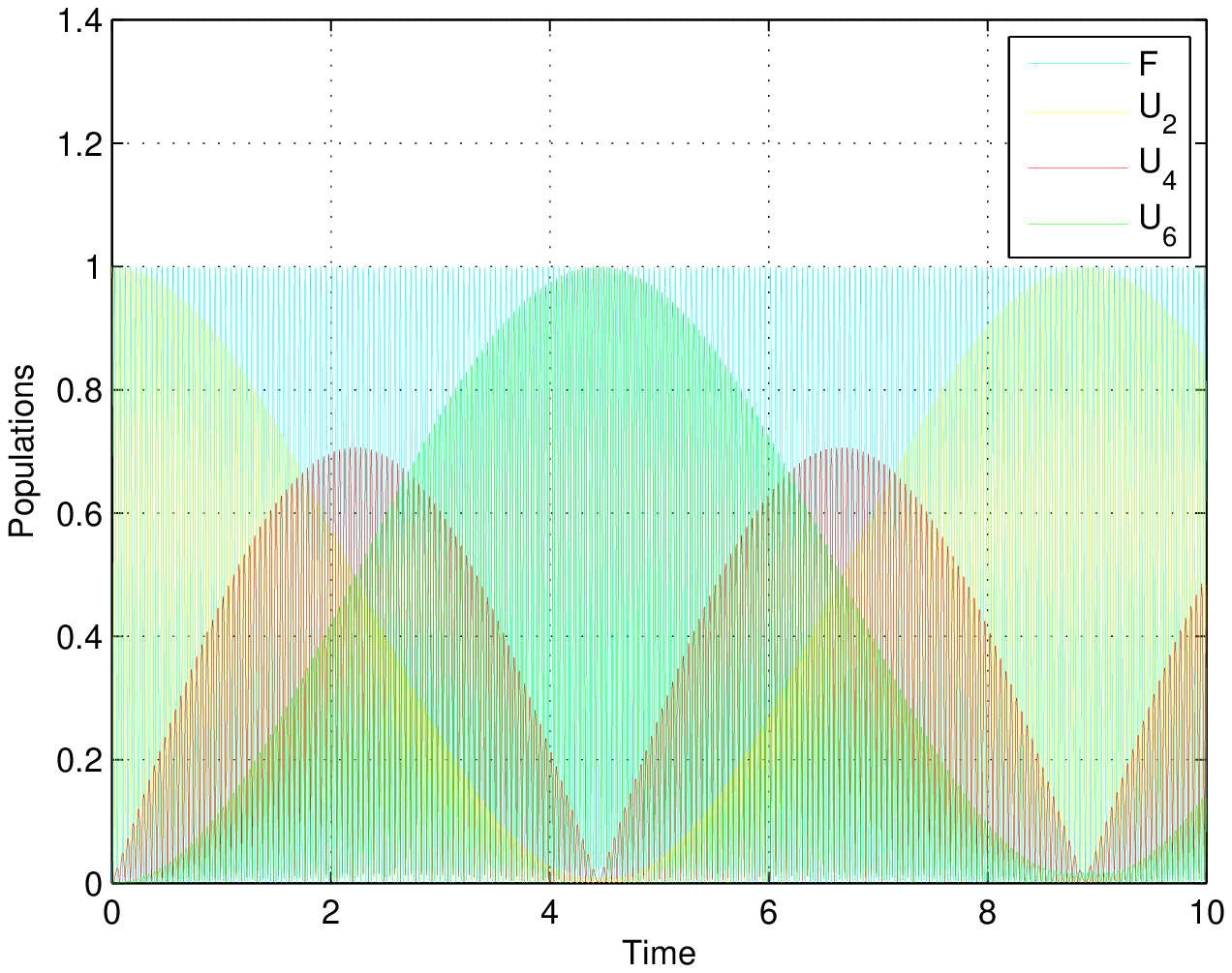}
\caption{} \label{Fig2}
\end{figure}
\newpage
\textbf{Figure Captions}
\itemize{}
\item Fig. 3. Perfect transfer of an arbitrary coherent state-based qubit (15),
at the nonresonance regime with $c=1$, $g=65$ and $\Delta=-600$ (in
units of $\omega$).
The blue curve represents the field mode of cavities, i.e. $F=|u_{_{b_{1}1}}(t)|^{2}+|u_{_{b_{1}3}}(t)|^{2}+|u_{_{b_{1}5}}(t)|^{2}$ with vanishing population. The populations of excitonic mode in QDs, i.e. $U_{2}=|u_{_{b_{1}2}}(t)|^{2}$, $U_{4}=|u_{_{b_{1}4}}(t)|^{2}$ and $U_{6}=|u_{_{b_{1}6}}(t)|^{2}$, are shown by yellow, red, and green curves respectively. At a time $t^{\ast}=195.4790$, perfect transfer of an arbitrary coherent state-based qubit from first QD to third one is satisfied which is equivalent to the expression $U_{6}=|u_{_{b_{1}6}}(t^{\ast})|^{2}=1$.
\begin{figure}
\centering
\includegraphics[width=445 pt]{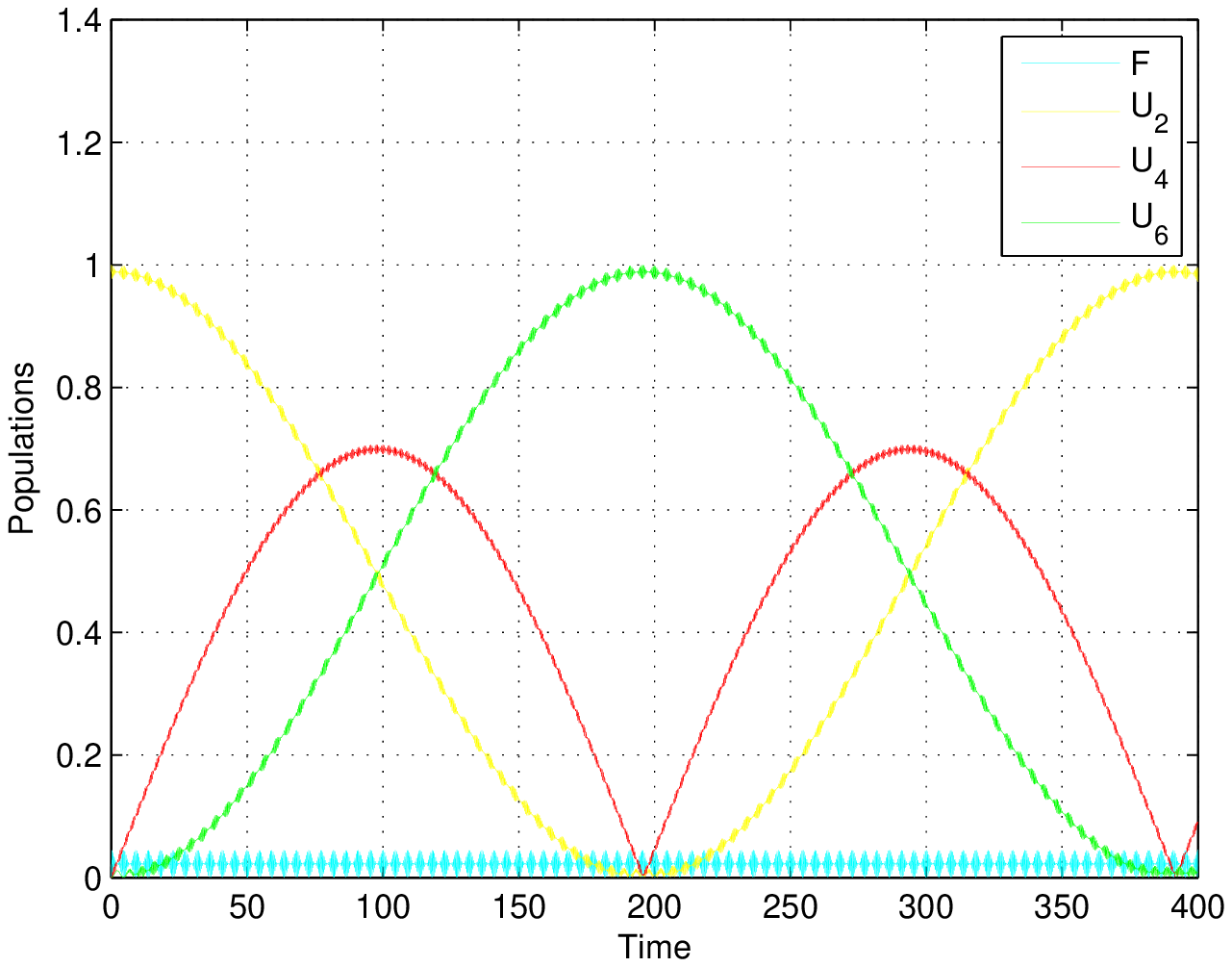}
\caption{} \label{Fig3}
\end{figure}
\end{document}